Using Astronomical Photographs to Investigate Misconceptions about Galaxies and Spectra:
Question Development for Clicker Use


Hyunju Lee[1,2,*] & Stephen E. Schneider[3]



Abstract

Many topics in introductory astronomy at the college or high-school level rely implicitly on using astronomical photographs and visual data in class. However, students bring many preconceptions to their understanding of these materials that ultimately lead to misconceptions, and the research about students' interpretation of astronomical images has been scarcely conducted. In this study we probed college students' understanding of astronomical photographs and visual data about galaxies and spectra, and developed a set of concept questions based on their common misconceptions. The study was conducted mainly in three successive surveys: (1) open-ended questions looking for students' ideas and common misconceptions; (2) combined multiple-choice and open-ended questions seeking to explore student reasoning and to improve concept questions for clickers; and (3) a finalized version of the concept questions used to investigate the strength of each misconception among the students in introductory astronomy courses. Here this study reports on the procedures and the development of the concept questions with the investigated common misconceptions about galaxies and spectra. We also provide the set of developed questions for teachers and instructors seeking to implement in their classes for the purpose of formative assessment with the use of classroom response systems. These questions would help them recognize the gap between their teaching and students' understanding, and ultimately improve teaching of the concepts.

---

[1]  Department of Teacher Education and Curriculum Studies, University of Massachusetts Amherst, MA 01002, USA
[2]  Leiden Observatory, Leiden University, Leiden, Netherlands (current)
*  hyunju.umass@gmail.com / hlee@strw.leidenuniv.nl
[3]  Department of Astronomy, University of Massachusetts Amherst, MA 01002, USA




Using Astronomical Photographs to Investigate Misconceptions about Galaxies and Spectra:
Question Development for Clicker Use

The Hubble Space Telescope (HST) has produced thousands of remarkable photographic images over the past 20 years. The HST images have had an important role of providing valuable scientific information to astrophysicists. In science classrooms, they are often used to attract students' attention to the beauty of the universe or to illustrate celestial phenomena, but students often have little guidance about how to examine astronomical images as data or how to interpret them. For example, students often gain the impression that galaxies are far more dense than they are in reality, they have little understanding of galaxy sizes and separations, and they misinterpret colors (sometimes false colors) in the images. In this study, we investigate how college students interpret photographic images of galaxies across the electromagnetic spectrum and we seek to understand their misconceptions. Based on the findings we developed concept questions that may be used with clickers in an introductory astronomy course to illuminate how students misinterpret astronomical images. Throughout this paper we use the term, *image*, as the meaning of visual data that contains physical information, often in the form of a photograph that is the common usage in the field of astronomy.

## Theoretical Framework

### Photographs and Misconceptions in Astronomy

Students bring ideas that have formed from their daily experiences to their learning



environment (Driver & Bell, 1986). This kind of prior idea is called a preconception, defined as "a

conception in a certain area that is present in a student prior to instruction" (Clement, 1993, p.

1241). Preconceptions often lead to misconceptions "that can conflict with currently accepted

scientific theory" (Clement, 1993, p. 1241). Many studies have investigated students'

misconceptions in science. There are several studies in astronomy as well, but they are relatively

limited in the topic areas they examine (Bailey & Slater, 2004). The astronomical concepts that

have been most studied are the shape of the earth and its gravity (Nussbaum & Novak, 1976), day-

night (Baxter, 1989; Dunlop, 2000; Vosniadou, 1994), cause of seasonal change (Atwood &

Atwood, 1996; Baxter, 1989; Dunlop, 2000; Schneps & Sadler, 1998), and phases of the moon

(Fanetti, 2001). Lelliott & Rollnick (2010) reviewed 103 peer-reviewed astronomy education

studies published between 1974 and 2008 and found that over 80% of the studies examined just

five topics: the earth, gravity, the day-night cycle, the seasons, and the earth-sun-moon system, and

only the remaining 20% of the studies examined other topics such as stars, the solar system, and

their sizes and distances. A few recent studies have begun to look at broader topics in astronomy

(Prather, Slater, Adams, & Brissenden, 2012) such the Big Bang theory (Prather, Slater, &

Offerdahl, 2002) or cosmology (Wallace, Prather, & Duncan, 2012), but misconceptions held

about many topics in astronomy are largely unexplored.



As students construct preconceptions from their daily experiences, they also form preconceptions from photographs and bring them to their science classes. While studies have long reported misconceptions in many science disciplines, only recently have they begun to investigate the potential role of image interpretation in forming these misconceptions. For example, Pozzer and Roth (2003) examined four categories of illustrations in high school biology textbooks identified according to whether or not there was a caption, relevant information, and/or further explanation. They then investigated how high school students interpreted photographs of each functional category (Pozzer & Roth, 2004). Another recent study used photographs to examine high school students' and pre-service teachers' understanding of Newton's 3[rd] law (Eshach, 2010). However, studies about photographs in science learning are relatively scarce compared to studies of other types of visual materials (Gilbert, 2008).

In professional scientific research, interpreting photographs is a fundamental and important procedure. This is especially true in astronomy, where photographs and spectra often provide our only clues to solving mysteries about astronomical objects and the universe. Striking, eye-catching photographs of stars or galaxies are commonly seen in daily life, and many students may have formed various preconceptions and misconceptions about astronomical concepts through photographs before learning about the subject in classes. What do they understand from astronomical photographs? How do they interpret them? Little is known about the answers to these



questions.

**Classroom Response Systems and Concept Questions**

Introductory astronomy classes at universities and colleges often have more than a

hundred students, making it challenging for an instructor to communicate with the students.

Instructors in such large classes often cannot help but do frontal teaching, simply conveying

information to their students. Educators' efforts have been aided over the last decade by

educational technologies such as *classroom response systems* (CRSs, or simply *clickers*). CRSs

allow an instructor to present simple numerical or multiple-choice questions to a class. Students

then enter their answers using the devices, and the system instantly collects their responses. The

students' responses are presented as a histogram, allowing the instructor and the students to see the

class's cumulative results. This technology can play various roles in class such as checking student

attendance or gathering student responses on exams, however, it is most valuable when it is used

by instructors to perform formative assessment and to understand students' various ideas in

classroom discussion (Feldman & Capobianco, 2008). In simple terms, instructors can decide by

looking at the histogram of responses whether they need to revisit the topic or can move to the next

one. Instructors can help students scaffold learning when they know whether their students

understand scientific concepts and what they are struggling with. In addition, CRS technology can

attract students into active participation, helping them overcome misconceptions by making them



conscious of their own background knowledge and preconceptions, by revealing gaps and

contradictions in students' understanding, and by identifying flaws in students' logic (Beatty,

2004).

To be successful in formative assessment with CRS technology, it is important to use

well-designed concept questions. Concept questions are a form of multiple-choice or simple

questions that are designed to explore students' understanding of scientific concepts (Beatty &

Gerace, 2009; Mazur, 1997). Mazur (1997) suggested in his book, *Peer Instruction,* that "Instead

of presenting the level of detail covered in the textbook or lecture notes, lectures consist of a

number of short presentations on key points, each followed by a *ConcepTest* – short conceptual

questions on the subject being discussed" (p. 10). He argued that those questions should satisfy the

following criteria: "focus on a single concept; not be solved by relying on equations; have

adequate multiple-choice answers; be unambiguously worded; and be neither too easy nor too

difficult" (p, 26). Good concept questions represent probable students' conceptions and reveal their

misconceptions (Mazur, 1997), so instructors can assess whether the students understand a

scientific concept and help them work through the issues that they misunderstand. Therefore, it is

necessary to determine student ideas and misconceptions in order to develop useful concept

questions.



Over the last several decades, concept questions have been developed in many science topics (e.g. Hestenes, Wells, & Swackhamer, 1992; Lightman & Sadler, 1993; Odom & Barrow, 1995; Russell, 1994; Treagust, 1988; Wandersee, Mintzes, & Novak, 1994). There are several sets of concept questions developed in astronomy as well, such as the *Astronomy Diagnostic Test* (Adams, Adrian, Brick, Brissenden, Deming, Hufnagel, Slater, & Zeilik, 1999), the *Light and Spectroscopy Concept Inventory* (Bardar, Prather, Brecher, & Slater, 2007), and the *Star Properties Concept Inventory* (Bailey, 2008). Notably, Green (2003) assembled approximately 400 questions over all general astronomy topics. However, those questions are mostly based on texts and/or diagrams, and they do not probe understanding of astronomical images. Astronomy is an observational science that gathers information mostly from light over a range of wavelengths. Therefore, visualization is an important starting point to understanding astronomical concepts. Green noted, "Visualization, with its emphasis on observation, is especially relevant for astronomy. The addition of graphics and figures to the ConcepTest Library is therefore an important goal for the future" (Green, 2003, p.36).

Our study investigates college students' interpretations of astronomical photographs and their misconceptions about galaxies and spectra, which are fundamental topics in introductory astronomy at the college and high-school level. Based on the findings of a series of surveys to uncover misconceptions, we developed multiple-choice concept questions that are applicable to



introductory astronomy courses or high school physics classes. In addition, the final version of

questions was surveyed with college students in two introductory astronomy courses to understand

how frequently students retained each misconception after instruction.

## Methods

### Context and Data Collection

This study was conducted primarily through three successive surveys. Survey 1 consisted

of 23 questions, all open-ended except for one multiple-choice question, exploring various ideas

about the photographs of galaxies and spectra. This survey was completed by college students in

an introductory astronomy course for non-majors in Spring 2011. The course covered general

topics in astronomy including stars, spectra, galaxies and modern astronomy. Because we wanted

to understand students' prior knowledge, Survey 1 was conducted early in the semester before

instruction on the topics of galaxies and spectra. Due to the length of the survey, it was split into

two parts and conducted in two different times on the course website; 117 students participated in

Survey 1 part 1, and 149 students in Survey 1 part 2. Students' responses on Survey 1 informed us

about variety of their misconceptions about galaxies and spectra.

Based on what was found from Survey 1, we developed a 2-tier concept survey instrument

that consisted of tentative multiple-choice concept questions and follow-up open-ended questions

(Survey 2). Students were asked to select their answer choices from among the multiple-choice



answers (selecting more than one answer choices was allowed), and then to write reasons for their answer choices in open-ended questions. Survey 2 was conducted with the same students as Survey 1 in the same class in late spring 2011 near the end of the semester, so we could learn what misconceptions they still held after instruction. Survey 2 was also divided into two parts due to its length; 198 students participated in Survey 2 part 1, and 162 students in Survey 2 part 2.

The questions were modified based on the information that we gained from the students' responses on Survey 2. Some of the questions were combined or removed and wordings were revised, and 41 multiple-choice concept questions were developed. Those questions were sent for review and to test *content validity* to two faculty members who were teaching introductory astronomy in two different universities. Content validity is a non-statistical type of validity study that estimates how well the items represent the specific intended domain of content. The faculty members reviewed the questions and provided comments on whether they covered topics appropriate for an introductory astronomy class and whether their wording was scientifically accurate. In addition, the faculty members were asked to identify any misconceptions they knew of relating to each question. Based on their comments, the questions were further modified.

As a result, we finalized Survey 3 with 38 concept questions. Survey 3 was used with students in two introductory astronomy classes at the end of fall 2011 (class A: 93 students completed) and at the end of spring 2012 (class B: 121 students completed). Survey 3 was used to



find out how commonly and consistently the misconceptions were held among the students in two

additional classes. Survey 3 was conducted through the course websites. The students were

allowed to choose only one answer choice in each multiple-choice question. All the surveys were

taken anonymously.

The participating students throughout the surveys were mostly freshmen (68.4%) and

sophomore (20.4%). Most students (80.9%) did not major in science. The introductory astronomy

course that they enrolled was their first astronomy class ever for most students (83.8%). Only a

small number of students (11.6%) had experience of learning astronomy in either middle or high

school, and even fewer (4.6%) had previously taken another astronomy course in college.

**Analysis**

The students' responses on each survey were analyzed in several ways. For example,

Survey 1 data were analyzed qualitatively. The students' responses were read thoroughly and

similar responses were grouped together. In that way, common themes of misconceptions were

identified for each scientific concept. On the other hand, Survey 2 data were analyzed using mixed

methods. The numerical data from multiple-choice questions were analyzed using descriptive

statistics and the strength of responses was measured for each answer choice. In the following

open-ended questions, students' common reasons for their answer choices were categorized. Based

on the results from Survey 2, we refined the concept questions to better represent students' various,



yet common and strong, misconceptions. Finally, *proportional similarity* was measured for Survey

3 data from the two introductory courses. This was done to understand how well the developed

concept questions represent students' common misconceptions. Proportional similarity, also called

*percentage similarity* (Huhta, 1979), is defined as:

$$PS\ (U,\ V) = \sum_{i=1}^{c} \min\ (f_{Ui},\ f_{Vi}) \qquad (\text{Renkonen, 1938}).$$

In simple terms, the proportional similarity is the sum of the lowest percent values between groups

divided by 100, and it tells the degree of similarity in the percentages of items (answer choices in

our study) between groups. It is a desirable method to measure similarity between distributions

(Vegelius, Janson, & Johansson, 1986).

## Results

### Student Ideas and Misconceptions

The findings reported in this section are based on students' open-ended responses to

Survey 1 and Survey 2. Here we present some of the most common misconceptions observed

among the students' various ideas about galaxies and spectra.

#### Radio Waves.

Students' interpretations of the images of the Milky Way made from observations at radio,

infrared and optical wavelengths (Figure 1) revealed various misconceptions about radio waves,



infrared emission, and the dust clouds in our galaxy. The question asked, "Why do you think the

colors in the radio image look much different from the ones in the infrared or optical images?"

Radio waves are longer wavelength light than optical and infrared light. While radio waves cannot

be seen with our eyes, the intensity and wavelength data recorded by radio detectors contain

similar information to that recorded in visible-wavelength images. Therefore, radio images such as

shown in Figure 1 are presented with *false* colors. Many students, however, thought that the radio

image looked different because radio waves are not light but *sound*. For example, "The radio

waves shows the images of the sounds of the milky way" (S11, survey1part1). Some students

thought that different wavelengths of light have different speeds, for example, "The radio image

might look different because the radio frequency is much different than infrared and optical where

it is much faster" (S54, survey1part1). This aligns with the well-known misconception that optical

light is thought to travel faster than radio waves (Zeilik, Schau, & Mattern, 1998; Lee, 2008). In

reality, the speed of light is the same for all wavelengths of electromagnetic radiation.

----------------------------- Insert Figure 1 around here -----------------------------

**Infrared.**

While radio waves are commonly thought of as *sound,* infrared is commonly thought of as

*heat*. The misconception here is more subtle in that many hot sources do emit primarily at infrared

wavelengths, but students' responses seemed to interpret infrared radiation as something quite

distinct from electromagnetic radiation and solely related to thermal emission. For example,



"Because the infrared image picks up on heat waves while the optical image is just what the milky way actually looks like" (S20, survey1part1), and "The radio picture is looking at the wavelengths. The infrared is looking for heat and energy. The optical image is looking at where there is a lot of stars together" (S16, survey1part1). This misconception may have come from students' familiarity with the use of infrared cameras that capture thermal images for night vision.

In general, the misconceptions about the Milky Way images at different wavelengths are summarized as: *radio shows its sound, infrared shows its heat and energy, and optical shows its light and reality,* as can be seen in the following responses: "The radio waves show the sound while the infrared shows a heat perspective while displaying the entire MW and the optical view shows a realistic view of the galaxy" (S82, survey1part1), and "One is measuring heat and the other is looking at light" (S101, survey1part1).

**Dust in the Milky Way.**

The following question was asked with regard to Figure 1, "Why do you think there is a bright horizontal band in the infrared image where the optical image mostly looks dark?" The dark horizontal band in the optical image is produced by dust clouds in our Galaxy. When starlight meets dust it is scattered and absorbed. At visible wavelengths the effect is so strong that a dark band is formed in the optical image across the region where the star density is actually highest. Because infrared radiation has a longer wavelength than visible light, the scattering is much



weaker. Therefore, infrared light passes through dust clouds and reaches our infrared cameras, allowing us to see the strong concentration of stars shown by the bright band in the infrared image.

Students' interpretations of the images revealed several misconceptions. Some believed that the dark features in the optical image were caused by *dark matter*, for example, "it may have to do with the amount of dark matter present, it is invisible to the naked eye however dark matter gives off a lot of energy" (S5, survey1part1). This appears to be a literal interpretation of the term *dark,* which astronomers and physicists use to mean *invisible* or *unseen* when referring to *dark matter,* and perhaps some confusion with the idea of *dark energy*. However, it is an interesting idea that dark regions in the optical image look bright in infrared image *because infrared cannot detect dark matter as optical does.*

Some students thought it was clouds blocking the view, but did not understand the nature or mechanism of obscuration by clouds. For example, "The infrared is picking up on light where as the optical doesn't have the same attention to detail, because there are clouds blocking the view" (S43, survey1part1). Some thought that *the band looked dark due to the mixture of various colors*. "The optical image shows the spectrum of visible colors that the eye can see. The dark image is due to a mixture of colors that make it look dark" (S11, survey1part1). This misconception appears to be linked to a preconception about mixing paint colors (subtractive color mixing) learned by children.



Another interesting misconception with regard to Figure 1 is that the images look different because each shows different *layers* of the Milky Way:

The structures of the images are different because they are each different layers of the galaxy. The optical image is the first layer, like the skin, of the galaxy. The infrared image is a layer below that, showing more of the innards of the galaxy. The radio image is the overall form of the galaxy, without major details or anything, only showing that the galaxy as a whole is in that shape (S14, survey1part1).

Some others thought that the images look different because of the direction from which the picture was taken, "Because of the angles and distances from where the pictures were taken" (S9, survey1part1); or it was taken at different places and different times, "The Milky Way looks different in the three images because they were photographed from different places at different times on Earth" (S25, survey1part1). These statements suggest that the students believe the Milky Way is so small that we have the ability to view it from different directions, rather than an object so vast that it is tens of millions of times larger than the greatest distance any space probes have traveled.

**Foreground Stars in a Galaxy Image.**

The Andromeda galaxy is the nearest spiral galaxy to us. Due to its relatively nearby distance, it covers a wide angular extent in the sky, and many foreground stars in the Milky Way



are also included in photographic images of the Andromeda galaxy (Figure 2). Foreground stars

are present in many astronomical photographs of galaxies or nebulae. Students were asked what

the thousands of tiny dots (noted as A in Figure 2) were and then to explain their reasoning. In

Survey 1, 85% students correctly thought that they were *stars*. Other popular responses were *dust*

or *distant galaxies*. However, among the students who correctly identified these objects as stars,

only a small fraction (15.7% in Survey 1) correctly responded that they belonged to our galaxy.

Rather, most students thought that they belonged to the Andromeda galaxy or were located

between our galaxy and the Andromeda galaxy.

----------------------------- Insert Figure 2 around here -----------------------------

Some students thought that the stars belonged to the Andromeda galaxy because "all stars

must belong to some galaxy and since Andromeda is the closest, they could belong to it" (S48,

survey1part1) or "because we can only take a picture of Andromeda, we cannot take a picture of

our galaxy because we are in it" (S1, survey1part1). Students also thought that the stars were

between our galaxy and the Andromeda galaxy, saying: "they [the stars] must be filling the empty

space in between galaxies" (S67, survey1part1), "stars are all over the universe including in the

areas between our galaxy and our next closest galaxy" (S108, survey1part1), or "they are just

hanging out in space, they don't look like they are caught up in the forces of the galaxies around

them" (S107, survey1part1). Even the ones who correctly responded that the stars belonged to our



galaxy in several cases offered spurious reasons, such as "it looks like Orion's belt which is in our galaxy" (S20, survey1part1), "the image of the galaxy looks similar to that of ours" (S21, survey1part1), or "because I think it [Andromeda galaxy] still is considered our galaxy" (S44, survey1part1). It is clear from these responses that many students do not recognize that most astronomical photographs necessarily contain completely unrelated objects in the foreground or background, and they try to construct explanations that connect these objects.

**Composition and Structure of a Spiral Galaxy.**

Students were shown a photograph of a spiral galaxy (Figure 3) and asked to identify several different types of regions and to explain the causes for the appearance of each region. Four kinds of regions in the galaxy were examined (labeled A, B, C, D in Figure 3): A is the bulge of the galaxy consisting largely of old stars; the dark features B are cool, dusty interstellar clouds where new stars form; the reddish blotches C are clouds of hydrogen gas where the radiation from recently-formed stars ionize the surrounding gas and causing it to glow; and the blue areas D are clusters of young stars.

The students provided various interpretations for each region. The central bright yellowish area A was commonly thought to be the *hottest place in the galaxy* having the most energy. This idea was probably based on noting the brightness of the region and associating the yellowish (or reddish) colors with the highest temperatures—another common misconception. Related to this



interpretation of a hot, energetic region, some students thought region A to be *the Sun, a massive star*, *the center of the solar system, dust or gas clouds* undergoing interactions, *the stellar nursery* where most new stars are born in the galaxy, or where *numerous stars are being sucked into a black hole*. The dark features B were thought to be *dark matter, black holes*, *void space*, *planets, or asteroids,* all objects that students expect do not emit much light, regardless of scale. The reddish blotches C were commonly thought to be *red giant stars* or *dying stars*, *very bright and hot stars*, *solar system*, *flames*, *heat waves*, or *smaller galaxies*. The pale blue areas D were thought to be *dust*, *gas clouds*, *water, oceans*, or *planets.* Most of these explanations seem to rely on passing familiarity with astronomical objects in relation to their colors, but their ideas are not scientifically accurate in terms of either the scale of objects in the image or a physical mechanism that could give rise to the appearance of these regions.

----------------------------- Insert Figure 3 around here -----------------------------

Students were additionally asked to explain the structure of a spiral galaxy. Most thought that the spiral structure was created *by a strong gravitational force pulling matter around it towards its center*, such as by *black holes*, for example, "Because the gravity of the black hole in the center of the galaxy is pulling everything toward it" (S96, survey1part1). While their ideas about gravitational attraction are partly on the right track, there seems to be an underlying notion that the overall spiral structure arises from a rapid swirling motion of matter into a central black



hole expressed most clearly here:

> Its image is a spiral so given whatever is in the middle of the structure, it is similar to that of a drain, it is sucking all things around it into its center creating the image shown (S5, survey1part1).

The student's reasoning seems to be recalled from their prior experience of seeing the spiral pattern in draining water, or perhaps even textbooks or simulations showing the spiraling pattern of gas outside a black hole. While it is correct that there is a massive black hole at the center of large galaxies, it is not the reason for the spiral arms which occur over a vastly larger scale. Spiral arms are generally produced by gravitational interactions between orbiting stars and interstellar clouds or with another galaxy.

### Elliptical Galaxies and Spiral Galaxies.

Students were shown images of the two major types of galaxies (Figure 4) and asked to explain the reasons for the difference of colors and shapes. Type A in the figure is an elliptical galaxy and looks yellow because it mostly consists of old stars. On the other hand, type B is a spiral galaxy and its spiral arms look blue because of hot, young stars. Although elliptical galaxies consist of mostly old stars, it does not mean that they are actually older than spiral galaxies. Rather, most elliptical galaxies have used up or lost their interstellar gas due to collisions with other galaxies, so they can no longer form new stars—which include the short-lived, very luminous blue



stars. Both elliptical and spiral galaxies are believed to have massive black holes at their centers,

although these are not detectable except through detailed analysis of orbital velocities near the very

center of the galaxy.

----------------------------- Insert Figure 4 around here -----------------------------

Several misconceptions were found about the different colors and structures of the

galaxies. One of the common misconceptions was that the galaxies looked different because they

were in *different stages of their life*. This is actually what many astronomers, including Edwin

Hubble, believed in the early 1900s when they began making some of the first deep photographs of

galaxies. Today the appearance of a galaxy and changes to it are understood to result in a complex

way from the history of interactions and mergers with other galaxies. The shape and type may

change as a result of these interactions.

Other common misconceptions about the difference in their appearance were that it was a

matter of *distance,* for example, type A is closer and type B is further; that it was a matter of the

*existence of a black hole at its center*, for example, type B has a black hole causing its spiral arms

but type A does not have it; or that it was because they were different objects, for example, type A

was actually a star while type B was a galaxy. Some thought that the images had different colors

because of the way the photograph was *taken: with different cameras, lenses, or filters*, or because

of *the angle* at which the galaxy was viewed.



Another interesting misconception was that the colors were interpreted as spectrally *red-shifted or blue-shifted,* with the type A galaxy moving away and the type B galaxy coming toward us:

The object type A is moving away from us causing a longer wavelength towards the red side of the color spectrum whereas in type B the object is moving towards us causing shorter wavelengths giving it the color of the blue side of the color spectrum (S71, survey1part2).

This confusion appears to have arisen as students grappled with the idea of Doppler shift, illustrating how students actively seek to construct an understanding of new ideas with their prior knowledge. As we will examine next, interpreting spectra also introduced a variety of other misconceptions.

**Spectra.**

It is essential to understand spectra in order to comprehend how astronomers derive information about elements, temperatures, distances and motions of galaxies. Students had a general understanding about the functions of spectra, but they showed a variety of misconceptions in their interpretation. For example, one question asked about the difference between spectra A and B in Figure 5. This question probes the concept of red-shift, with the spectral lines shifted to longer (redder) wavelengths because the more-distant galaxy (B) is moving away from us faster.



However, many students incorrectly thought *the spectra indicated that the objects had different temperature*, *contained different elements*, or *had different ages*.

----------------------------- Insert Figure 5 around here -----------------------------

**Distances of Galaxies based upon Spectra.**

An image of several hypothetical galaxy spectra (Figure 6) was given to students with the information that the spectra were obtained from four different galaxies. The students were asked whether or not they could arrange the distances of the galaxies from us with the information seen in the spectra, and why or why not. The astronomical explanation is that the farther the galaxy, the faster it is moving away from us, therefore the more its spectral lines are red-shifted. So the galaxy that has the least red-shift to its lines is the closest to us and the galaxy that has the most red-shift is the farthest from us. Therefore, the correct answer is B-C-D-A, from closest to farthest.

----------------------------- Insert Figure 6 around here -----------------------------

Students who responded that the distances could *not* be arranged with the given information mostly thought that *spectra were only used to understand an object's composition*, for example, "No, these are to tell what kind of materials are in them, not distance" (S23, survey1part2).

Although many students responded that the spectra could be used to tell the relative distances of galaxies and correctly arranged the distance, their reasoning showed several misconceptions in the interpretation of the spectra. For example, they thought that *the number of*



*absorption lines and their widths are related to the galaxy's distance*: "BCDA, more black lines and wider they are, the further away they are" (S98, survey1part2). In addition, many students seemed to correctly arrange the distances by simple guesswork, choosing *the one that started the line closest to the left edge of the spectrum* and so on, "BCDA, when the lines start to closer to the beginning of the line spectra the closer the galaxy" (S118, survey1part2). A few students referred to *brightness of the spectra* in considering the distance to the galaxies, "I'm guessing that you can tell how far they are by seeing their brightness from these spectra" (S49, survey1part2); or *number of absorption lines in the red part of spectra*, "the closer the galaxy the hotter it will be and the more absorption would occur in the red zone" (S50, survey1part2).

**Temperatures from Spectra.**

To further understand the students' interpretation of spectra, spectra and wavelength-intensity graphs of three different stars (Figure 7) were presented with a question asking about their relative temperature. In nature, the hotter the star, the higher the intensity it will have at shorter wavelength (brighter at the blue end of a visible light spectrum), therefore A is the hottest while B is the coolest.

Some students thought that the star C had the highest temperature *because it has a high intensity most uniformly across the spectrum.* Some students thought that the star B was the hottest *because it has more fluctuations on the intensity graph representing more activity*, *because it has*



*more absorption lines in certain parts of the spectrum* than the others, or *because its intensity graph is higher in the red part of the spectrum.* The last one concurs with the previously reported misconception, which students believed blue as indicating low temperature and red as indicating high temperature (Cid, Lopez, & Lazarus, 2009).

----------------------------- Insert Figure 7 around here -----------------------------

**Concept Questions**

So far, we have reported misconceptions found in the Survey 1 and the Survey 2 that were common to the students. Additional misconceptions were found in the surveys that were unique to one or a small number of students. The questions and their answer choices on the Survey 3 were all constructed from the students' responses from Survey 1 and Survey 2. In other words, misconceptions that were commonly found among the students became the choices of the concept questions.

----------------------------- Insert Table 1 around here -----------------------------

The questions basically ask about 12 concepts relating to galaxies and spectra (Table 1). The mean scores of the students' performance on the Survey 3 were 12.4 for the Fall 2011 class and 9.2 for the Spring 2012 class. The distribution of their scores showed a similar bell-shape with peaked at the range of 6~10 between the two classes (Figure 8). Although the students took the survey at the end of the semester after the instructions, their performances were mostly in the



range of low scores. Considering the pure guessing score of the survey is 8.9 that would be achieved purely by chance, their performance on the survey is generally quite low.

Popular misconceptions were still observed among many students and their tendency was similar between the two classes. Among the 38 questions, 30 questions had the proportional similarity larger than 0.8 between the two classes where the survey was administered. The lowest proportional similarity was 0.691 and it was the only question that was lower than 0.7. On average for all 38 questions, the proportional similarity between the two classes was 0.844, which indicates that the percentages of students who chose certain misconception are similar in both classes, even though the students were drawn from different classes taught by different instructors. The similarity of incorrect answers between the two classes suggests that they are founded on common preconceptions that lead to persistent misconceptions and these misconceptions may be generally common among students even after taking an introductory astronomy course.

----------------------------- Insert Figure 8 around here -----------------------------

## Conclusions

This study investigated college students' misconceptions about galaxies and spectra with the use of astronomical photographs and spectral images and developed a set of multiple-choice concept questions. Some examples of misconceptions found in this study are: radio waves are sound, and travel slower than optical light; infrared means heat; the horizontal band of the Milky



Way looks dark in the optical image due to its dark matter; the spiral structure of a spiral galaxy is due to a strong gravitational force at the center, and it is created by a black hole; and different colors between elliptical galaxies and spiral galaxies are because of one red-shifted and the other blue-shifted. Students also believed that the foreground stars taken in the photograph of Andromeda galaxy belonged to the Andromeda, not to our galaxy; misinterpreted colors of objects and the compositions of a galaxy on the photograph, and misunderstood the information that spectral images represent in regard to its absorption lines and strength. Based on the findings we developed 38 concept questions that their answer choices were all constructed from the common misconceptions and that would be useful in introductory astronomy courses.

## Discussion and Implications of the Study

In general, introductory astronomy courses are popular and they are often large with more than a hundred students. Very few of the students we surveyed had received instruction in astronomy in middle or high school, and for most students the introductory course at college was their first astronomy class. Therefore, they had not had much opportunity to learn astronomy in a formal setting, although they certainly had seen images of galaxies and spectra and had developed some preconceptions before taking the course. Even after instruction, many preconceptions persisted as misconceptions, and this may reflect that the students had little opportunity to understand the context of objects that they were presented in such a way on the photographs.



However, it is still not clear *why* students created such misconceptions and it remains as our future work.

Misconceptions, when instructors recognize them in a timely manner, can be a good starting point for organizing instructional approaches to help overcome students' confusion and provide them with proper foundations. Due to the high number of students in many introductory astronomy courses, however, it is often difficult to recognize and address students' preconceptions. Moreover, little has been studied about misconceptions about more advanced topics such as galaxies and spectra. Therefore, the concept questions constructed from common misconceptions found in this study should prove useful in introductory astronomy courses that explore these topics. This type of question is not only helpful for instructors to recognize students' misconceptions, but they also can be used with CRSs or other techniques to directly address the points of confusion during class and to help the students construct a better foundational understanding. Therefore it is more useful to use these questions for the purpose of formative assessment than of summative assessment that simply checks students' knowledge. It is desirable that instructors present the questions, give students to think as a small group (or individually), and receive their answer choices with their own explanations. Then, instructors facilitate whole classroom discussion that students can share their various ideas. In that way instructors have opportunities to hear students' reasoning and find the gap between their teaching and the students' understanding. This also helps



students learn from different perspectives and develop their thinking to become sound reasoning. This can be done without the use of CRSs, but it is certainly a good assist for instructors to manage classroom discussion as well as for students to be much engaged in discussion when using CRSs throughout the process.

Developing good concept questions is challenging (Lee, Feldman, & Beatty, 2012), but they are useful when employed with CRSs for the purpose of formative assessment in classroom discussion. CRSs are particularly useful in a large class because otherwise it is difficult to engage every student. This paper suggests one way of developing concept questions by investigating common misconceptions among students in a large class by: firstly, gathering open-ended written responses from the students; secondly, categorizing them into several themes of misconceptions and measuring the strength of each misconception with 2-tier questions that consist of multiple-choice and follow-up open-ended questions that do not limit students to choose only one answer; and finally modifying and *tightening* the concept questions to be as scientifically rigorous as possible while remaining appropriate for the level of the course by having reviews from colleagues who are experienced at teaching introductory astronomy.

Finally, we offer some remarks about the use of photographs and visual data in science education. This study showed that a variety of misconceptions were identified in the interpretation of photographs. In astronomy, photographs contain important data for scientific research, so



helping students to analyze and interpret visual data is a critical part of their education. At the same time, it is clear that students approach scientific images with a variety of preconceived ideas about how to interpret them. Therefore, it is essential that we understand how students interpret these visual data so we can find better ways of providing instruction to improve students' learning in astronomy. Photographs and their interpretation should be examined more thoroughly in studies of student learning in science classes, especially in astronomy.

See Supplemental Material at

[http://journals.aps.org/prstper/abstract/10.1103/PhysRevSTPER.11.020101#supplemental] for the set of developed concept questions.

## Acknowledgement

This work is supported by NASA HST Education and Public Outreach grant: HST-EO-12198.04. The opinions expressed are solely those of the authors and no official endorsement from the funder should be presumed. We thank Prof. Daniela Calzetti and Prof. Jessica Rosenberg for their reviews and comments about the concept questions. We also thank Prof. Kathleen Davis for her advice in research. Most of all, we thank the participating college students.

**Figures**

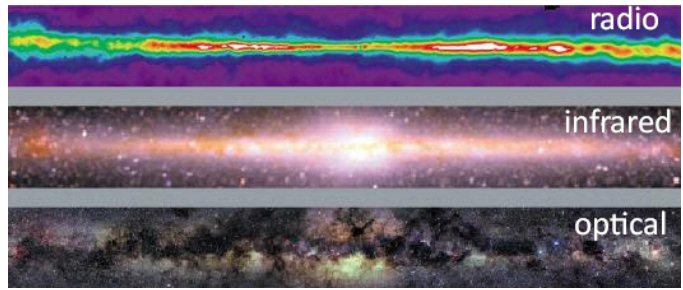

Figure 1. The Milky Way observed at radio, infrared, and optical

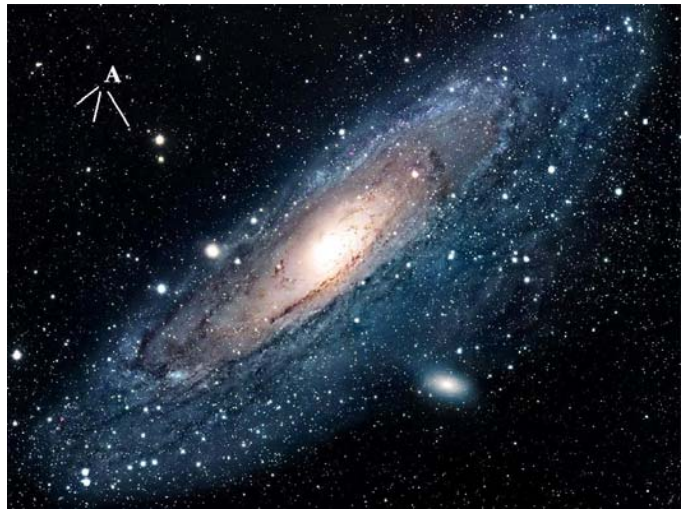

Figure 2. The Andromeda Galaxy

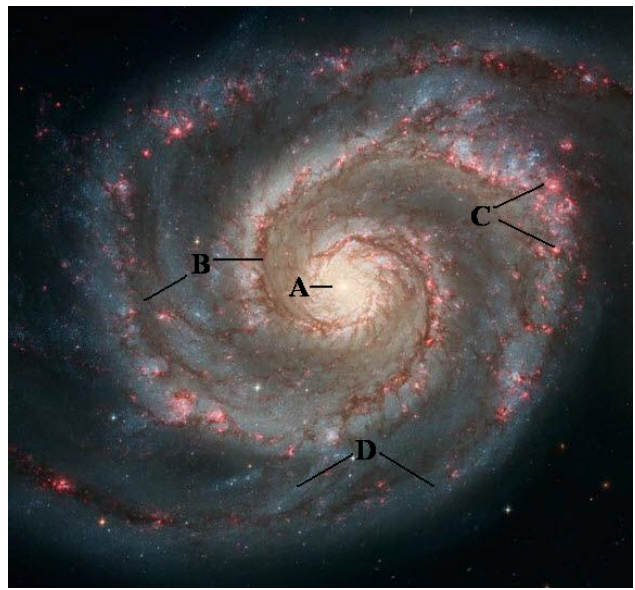



Figure 3. Spiral Galaxy

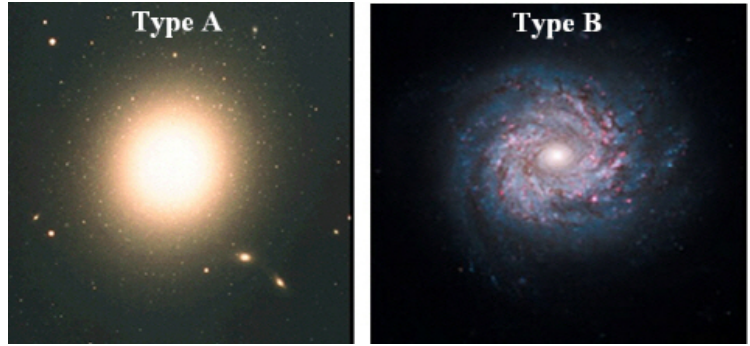

Figure 4. Elliptical galaxy (type A) and Spiral galaxy (type B)

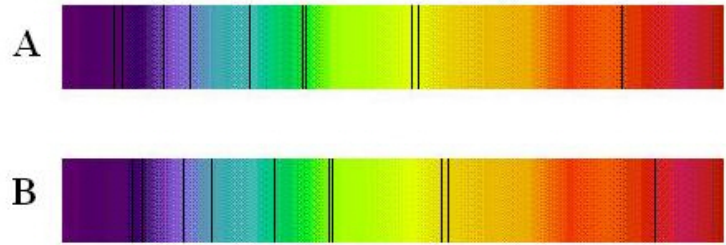

Figure 5. Spectra: Red-shift

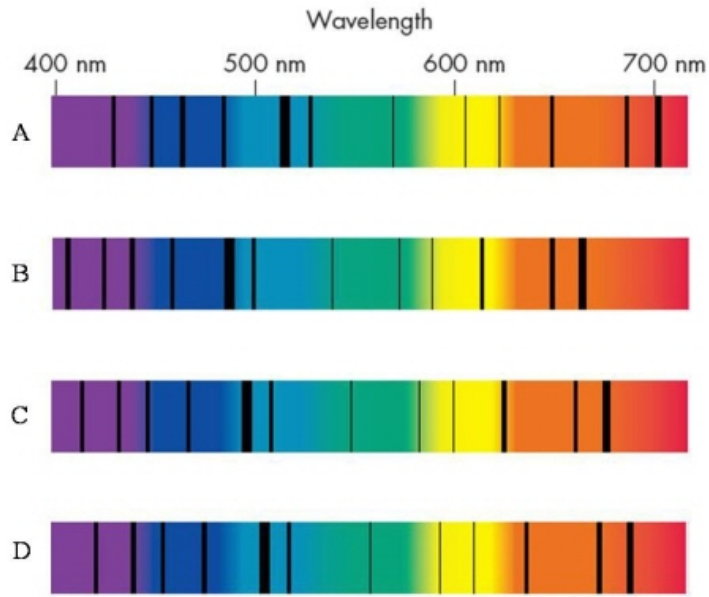

Figure 6. Spectra: Distance of galaxies



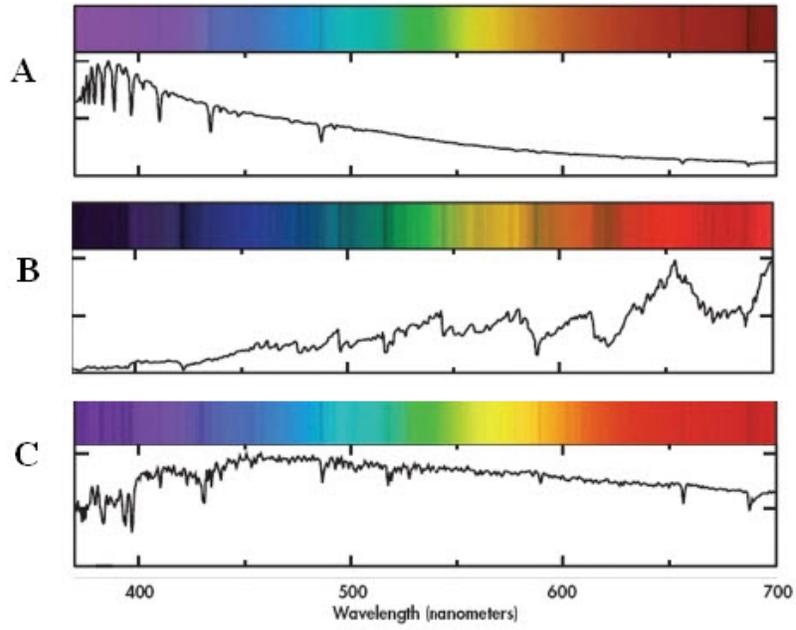

Figure 7. Spectra: Stellar temperature

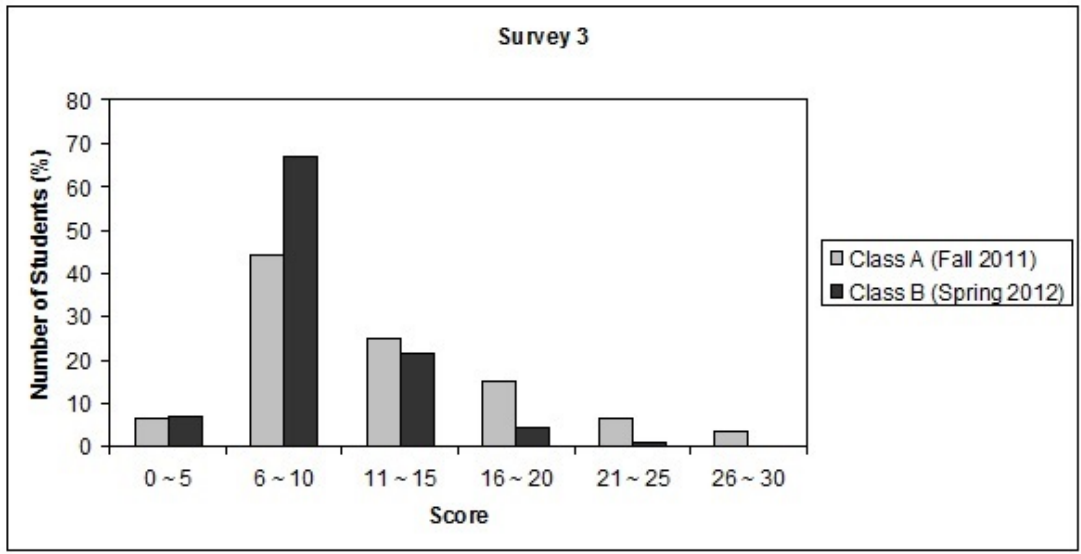

Figure 8. Students' scores on the Survey 3



**Tables**

| Table 1. Topics of concept questions | |
|---|---|
| Question Number | Concepts |
| Q1, Q2, Q3, Q4 | Fundamental characteristics of light |
| Q5, Q6, Q7, Q8, Q13 | Radio waves and Infrared |
| Q9, Q10, Q11, Q12 | Compositions of a Spiral galaxy |
| Q14 | Foreground stars |
| Q15, Q16, Q17, Q18, Q19, Q20 | Galaxies on the Hubble Deep Field image |
| Q21,Q22, Q23, Q24 | Elliptical vs. Spiral galaxies |
| Q25, Q26, Q27 | Composition of spiral arms and the movement of disk stars of a spiral galaxy |
| Q28, Q29, Q34 | Spectra: Intensity |
| Q30, Q31 | Spectra: Broadening effect |
| Q32, Q33, Q35 | Spectra: Distance of galaxies & Hubble's law |
| Q36, Q37 | Milky Way observed on earth |
| Q38 | Gravitational lensing |